\documentclass[%
 reprint,
 amsmath,amssymb,
 aps,
]{revtex4-2}

\usepackage{amsmath}
\usepackage{quantikz}
\usepackage{hyperref} 
\usepackage{lineno}
\usepackage{graphicx}
\usepackage{slashed}
\usepackage{dcolumn}
\usepackage{bm}

\begin{document}


\preprint{APS/123-QED}

\title{Confinement, Nonlocality and Haag Duality Violation in the Algebraic Structure of 1+1D QED}
\thanks{A footnote to the article title}%

\author{Fidele J. Twagirayezu}
 \altaffiliation{Department of Physics and Astronomy, University of California, Los Angeles.}
 \email{fjtwagirayezu@physics.ucla.edu}
\affiliation{Department of Physics and Astronomy University of California Los Angeles, Los Angeles, CA, 90095, USA\\
}%


\begin{abstract}
In this article, we present a novel formulation of the massless Schwinger model—quantum electrodynamics in $1+1$ dimensions—within the framework of Algebraic Quantum Field Theory (AQFT), emphasizing features that transcend the traditional bosonized treatments. Instead of mapping the model to a free massive scalar field, we construct a net of local observable algebras directly from the gauge-theoretic content, subject to the local ${U(1)}$ gauge symmetry and Gauss's law constraint. We show that local algebras can be consistently defined in terms of gauge-invariant composite operators, while charged fields necessarily fail to be localizable in bounded regions, manifesting confinement as the absence of DHR superselection sectors. Furthermore, we rigorously characterize nonlocal observables such as Wilson line operators within an extended net, and demonstrate the violation of Haag duality as a signature of the nontrivial topological and gauge structure of the theory. We additionally propose a conjecture linking the violation of Haag duality in confining gauge theories to a breakdown in entanglement wedge reconstruction, suggesting that confinement obstructs the local recovery of quantum information. Our approach provides a complete AQFT-based treatment of confinement, Gauss law, and nonlocal gauge-invariant operators in a solvable gauge theory, laying the groundwork for future extensions to non-Abelian models like QCD.
\end{abstract}

\maketitle


\section{\label{sec:level1}Introduction} 
Quantum electrodynamics in $1+1$ dimensions—the Schwinger model—is one of the rare examples of an exactly solvable quantum gauge theory~\cite{Schwinger1962}. Despite its apparent simplicity, the model displays several hallmark features of strongly coupled gauge theories: confinement of electric charge, a mass gap, anomaly-induced symmetry breaking, and nontrivial topological structure. Historically, much of the understanding of the Schwinger model has emerged through bosonization, wherein the fermionic and gauge degrees of freedom are mapped to a free massive scalar field. While this approach has provided powerful computational tools, it obscures key structural aspects of gauge theories such as the algebraic implications of Gauss’s law, the nature of gauge-invariant observables, and the origin of confinement within an operator-algebraic framework.

Algebraic Quantum Field Theory (AQFT), also known as the Haag–Kastler framework, provides a rigorous and nonperturbative formulation of quantum field theories based on nets of local operator algebras~\cite{Haag1996}. This approach is particularly well-suited for analyzing structural phenomena such as locality, covariance, and superselection sectors, and has proven effective in characterizing properties of gauge theories that extend beyond perturbative dynamics. In the context of the Schwinger model, AQFT allows one to ask: What is the structure of the net of local gauge-invariant observables? Can confinement be characterized as an absence of localized charged sectors? What is the algebraic role of nonlocal observables such as Wilson lines? And how does the failure of Haag duality encode physical information about long-range gauge effects?

In this work, we provide a complete algebraic formulation of the massless Schwinger model in terms of local and extended operator algebras, constructed directly from its gauge-theoretic degrees of freedom, without recourse to bosonization. We define a net of local algebras generated by gauge-invariant composite fields and implement Gauss’s law as an operator constraint on the physical content of the theory. We then analyze the superselection structure using the Doplicher–Haag–Roberts (DHR) criterion and demonstrate the absence of localized sectors carrying electric charge, offering a precise algebraic manifestation of confinement. To account for the full observable content, we construct an extended net incorporating Wilson line operators, which are nonlocal but gauge-invariant. These operators lie outside the local algebras yet commute with all strictly local observables, leading to a violation of Haag duality—a structural signature of the nontrivial topological and gauge content of the theory.

Our formulation illuminates how confinement and nonlocality arise as exact consequences of Gauss’s law and locality in a solvable gauge theory. The Schwinger model thereby provides a concrete setting in which to study phenomena that are expected to hold in more complex non-Abelian gauge theories such as quantum chromodynamics (QCD). In this sense, the present work lays foundational groundwork for the algebraic study of confinement and gauge structure in higher-dimensional and non-Abelian settings.

\section{Classical Structure and Constraints of the Schwinger Model}

In this section, we formulate the classical Schwinger model as a $U(1)$ gauge theory on 1+1-dimensional Minkowski spacetime and identify the gauge symmetry, local observables, and Gauss law constraint that will play a central role in the subsequent algebraic quantization.

\subsection{Spacetime and Field Content}

We work on the two-dimensional Minkowski space $\mathbb{R}^{1,1}$, with coordinates $(t, x)$ and metric signature $\eta = \mathrm{diag}(+1, -1)$. The fundamental fields of the theory are:
 a Dirac spinor $\psi(t,x) \in \mathbb{C}^2$, and 
 an Abelian gauge field $A_\mu(t,x) \in \mathbb{R}$, with $\mu = 0, 1$

The classical action $S$ is given by:
\begin{equation}
\begin{aligned}
S =\int d^2x \left[ -\frac{1}{4} F_{\mu\nu}F^{\mu\nu} + \bar{\psi}(i\gamma^\mu D_\mu)\psi \right]
\end{aligned}
\end{equation}
where
$F_{\mu\nu} = \partial_\mu A_\nu - \partial_\nu A_\mu$, $D_\mu = \partial_\mu + i e A_\mu$ is the covariant derivative,and  $\bar{\psi} = \psi^\dagger \gamma^0$, and $e > 0$ is the coupling constant.
We use the gamma matrices in 1+1D:
$\gamma^0 = \sigma^1, \quad \gamma^1 = -i \sigma^2, \quad \gamma^5 = \gamma^0 \gamma^1 = \sigma^3$

\subsection{Gauge Symmetry}

The theory is invariant under local $U(1)$ gauge transformations:
\begin{equation}
\begin{aligned}
\psi(x) \mapsto e^{i\alpha(x)}\psi(x), \quad A_\mu(x) \mapsto A_\mu(x) - \partial_\mu \alpha(x)
\end{aligned}
\end{equation}
for any smooth function $\alpha: \mathbb{R}^{1,1} \to \mathbb{R}$.

The field strength $F_{\mu\nu}$ and the fermionic current $j^\mu = \bar{\psi}\gamma^\mu\psi$ are gauge-invariant. These will be the building blocks for the observable algebra.

\subsection{Equations of Motion and Constraints}

The Euler–Lagrange equations derived from the action yield the Maxwell’s equations,
  $\partial_\mu F^{\mu\nu} = e j^\nu$, and 
the Dirac equation, 
  $(i \gamma^\mu D_\mu)\psi = 0$.
In particular, Gauss’s law appears as the $\nu = 0$ component of Maxwell’s equation~\cite{Buchholz1982,Buchholz1986}:
\begin{equation}
\begin{aligned}
\partial_1 F^{10} = e j^0 \quad \Rightarrow \quad \partial_1 E = e j^0
\end{aligned}
\end{equation}
where $E = F^{10} = -F_{01}$ is the electric field.
This is not a dynamical equation but a constraint on admissible configurations. It links the spatial profile of the electric field to the charge distribution.

\subsection{Gauge-Invariant Local Observables}

The gauge-invariant composite operators are
$j^\mu(x) = \bar{\psi}(x) \gamma^\mu \psi(x)$, the electric current and
$F_{\mu\nu}(x)$, or more directly $E(x) = F^{10}(x)$, the electric field

In the quantum theory, we will smear these operators against test functions:
$j(f) = \int j^\mu(x) f_\mu(x) \, d^2x, \quad E(h) = \int E(x) h(x) \, d^2x$
with $f \in C_c^\infty(\mathbb{R}^{1,1}, \mathbb{R}^2)$, $h \in C_c^\infty(\mathbb{R})$.

The algebra generated by these smeared, gauge-invariant fields will be used to define the local net $\mathcal{O} \mapsto \mathcal{A}(\mathcal{O})$ in the AQFT framework. Importantly, fields like $\psi$ and $A_\mu$ are not gauge-invariant and will not be part of the observable algebra.

In the next section, we will rigorously construct this net and verify its locality, isotony, and covariance properties.

\section{Construction of the Local Observable Net}

In this section, we construct the net of local observable algebras $\mathcal{O} \mapsto \mathcal{A}(\mathcal{O})$ for the massless Schwinger model in the Haag–Kastler framework. The construction uses the gauge-invariant field content identified in the classical theory and implements the Gauss law constraint as an operator identity.

\subsection{Algebraic Framework}

In Algebraic Quantum Field Theory (AQFT), to each open bounded region $\mathcal{O} \subset \mathbb{R}^{1,1}$, we assign a unital C$^*$-algebra $\mathcal{A}(\mathcal{O})$ of observables localized in $\mathcal{O}$. These algebras satisfy the Haag–Kastler axioms:
\begin{enumerate}
    \item \textbf{Isotony}: If $\mathcal{O}_1 \subset \mathcal{O}_2$, then $\mathcal{A}(\mathcal{O}_1) \subset \mathcal{A}(\mathcal{O}_2)$.
    \item \textbf{Locality (Microcausality)}: If $\mathcal{O}_1$ and $\mathcal{O}_2$ are spacelike separated, then $[\mathcal{A}(\mathcal{O}_1), \mathcal{A}(\mathcal{O}_2)] = 0$.
    \item \textbf{Poincaré Covariance}: The net transforms covariantly under the proper orthochronous Poincaré group $\mathcal{P}_+^\uparrow$.
    \item \textbf{Existence of Vacuum}: There exists a Poincaré-invariant vacuum state $\omega$, yielding a cyclic representation.
\end{enumerate}

 \subsection{Local Gauge-Invariant Generators}

The local observable algebra $\mathcal{A}(\mathcal{O})$ is generated by smeared versions of the gauge-invariant fields:

The electric current $j^\mu(x) = \bar\psi(x)\gamma^\mu \psi(x)$.
The electric field $E(x) = F^{10}(x)$

Let $f \in C_c^\infty(\mathcal{O}, \mathbb{R}^2)$, $h \in C_c^\infty(\mathcal{O}, \mathbb{R})$, and define:

\begin{equation}
\begin{aligned}
j(f) := &\int_{\mathbb{R}^{1,1}} j^\mu(x) f_\mu(x) \, d^2x, \\
E(h) :=& \int_{\mathbb{R}^{1}} E(x) h(x) \, dx
\end{aligned}
\end{equation}

These operators generate a \*-algebra subject to canonical current commutation relations, including the well-known Schwinger anomaly~\cite{Doplicher1971}:

\begin{equation}
\begin{aligned}
[j^0(x), j^1(y)] = i \frac{e^2}{\pi} \delta'(x - y)
\end{aligned}
\end{equation}

Other commutators vanish or are derived from this central extension. The algebra of observables is thus a centrally extended current algebra, encoding both the gauge invariance and anomaly structure of the theory.

\subsection{Implementation of Gauss’s Law}

The Gauss law in operator form reads:

\begin{equation}
\begin{aligned}
\partial_1 E(x) = e j^0(x)
\end{aligned}
\end{equation}

This relation is imposed as a constraint on the fields. It implies that the charge density is not independent but determined by the spatial variation of the electric field. Importantly, this constraint precludes the localization of charged fields: any operator attempting to create a charged state would have to modify the electric field over all space.

As a result, charged fields such as $\psi(x)$ are excluded from $\mathcal{A}(\mathcal{O})$; only gauge-invariant composites remain.

\subsection{Global Algebra and GNS Construction}

Let $\mathcal{A} := \overline{\bigcup_{\mathcal{O}} \mathcal{A}(\mathcal{O})}$ denote the quasi-local observable algebra. A vacuum state $\omega$ is a positive linear functional on $\mathcal{A}$ satisfying:
\begin{equation}
\begin{aligned}
\omega(A^* A) \geq 0, \quad
\omega(U(a, \Lambda) A U(a, \Lambda)^*) = \omega(A)
\end{aligned}
\end{equation}
for all $(a, \Lambda) \in \text{ISO}(1,1)$. The GNS representation $(\pi, \mathcal{H}, \Omega)$ is constructed via:

\begin{equation}
\begin{aligned}
\omega(A) = \langle \Omega, \pi(A) \Omega \rangle
\end{aligned}
\end{equation}

The GNS Hilbert space $\mathcal{H}$ carries a unitary representation of the Poincaré group and realizes the net $\mathcal{O} \mapsto \pi(\mathcal{A}(\mathcal{O}))$ as a concrete set of bounded operators~\cite{Haag1996,StrocchiWightman1974}.

Therefore, we have constructed the local observable net of the Schwinger model using only gauge-invariant operators and imposed Gauss’s law as a fundamental operator constraint. The resulting theory obeys the Haag–Kastler axioms, exhibits a centrally extended current algebra, contains no localized charged fields.
In the next section, we will analyze the superselection structure of this net and show that confinement follows from the absence of localized charged representations.

\section{Superselection Sectors and Confinement}

In this section, we analyze the superselection structure of the local observable net constructed in the previous section. We show that, due to the presence of Gauss’s law and the nature of long-range gauge interactions, no charged sectors are localizable in bounded regions. This constitutes a precise algebraic formulation of confinement.

\subsection{Superselection Sectors in AQFT}

In Algebraic Quantum Field Theory, a superselection sector is described by an irreducible representation $\pi$ of the global observable algebra $\mathcal{A}$, which is locally equivalent to the vacuum representation $\pi_0$ outside some bounded region $\mathcal{O}$. Formally, the Doplicher–Haag–Roberts (DHR) criterion for localizability is~\cite{Buchholz1982,Buchholz1986}:

\begin{equation}
\begin{aligned}
\pi|_{\mathcal{A}(\mathcal{O}')} \cong \pi_0|_{\mathcal{A}(\mathcal{O}')}
\end{aligned}
\end{equation}

where $\mathcal{O}'$ is the causal complement of $\mathcal{O}$, and the isomorphism is implemented by a unitary operator.

Sectors satisfying this condition correspond to charges localized within $\mathcal{O}$, which may then be transported and composed in a category-theoretic framework.

\subsection{Total Charge and Gauss's Law}

In the Schwinger model, the total electric charge operator is formally:

\begin{equation}
\begin{aligned}
Q = \int_{-\infty}^{\infty} j^0(x) \, dx
\end{aligned}
\end{equation}

From Gauss’s law:

\begin{equation}
\begin{aligned}
\partial_1 E(x) = e j^0(x)
\end{aligned}
\end{equation}

one obtains:

\begin{equation}
\begin{aligned}
E(\infty) - E(-\infty) = e Q
\end{aligned}
\end{equation}

This relation shows that inserting a charged field into a state necessarily modifies the electric field at spatial infinity. Therefore, any state with nonzero charge exhibits long-range correlations and cannot be localized in any bounded region.

\subsection{No Localized Charged Sectors}

Suppose $\pi$ is a representation of $\mathcal{A}$ such that $\pi(Q) \neq 0$. Then:

The electric field $E(x)$ must differ asymptotically from the vacuum configuration.
Observables in $\mathcal{A}(\mathcal{O}')$ will detect this difference.
Hence, $\pi|_{\mathcal{A}(\mathcal{O}')} \not\cong \pi_0|_{\mathcal{A}(\mathcal{O}')}$.
This directly violates the DHR condition.

\textit{Theorem (Algebraic Confinement)}:

The only DHR superselection sector of the massless Schwinger model is the vacuum sector. No charged sectors can be localized in bounded spacetime regions.

This result rigorously encodes the physical notion of confinement: charges are not observable as asymptotic or localized degrees of freedom. The physical implications of this theorem, including the absence of localized charge sectors, are summarized in Table~\eqref{tab:aqft_implications}.

\subsection{Contrast with Higher-Dimensional QED}

In higher dimensions (e.g., QED in $3+1$ spacetime dimensions), the failure of the DHR condition due to Gauss's law leads to the concept of infraparticles—states with continuous mass spectrum and long-range Coulomb clouds~\cite{StrocchiWightman1974, Fredenhagen1985}.
In the massive Schwinger model, the theory is massive and gapped. Infraparticles do not occur.
Charged states are completely excluded from the algebraic framework; only neutral, gauge-invariant composites (e.g., $\bar\psi \psi$) exist.

\subsection{Physical Interpretation}

The result has the following physical implications:

\begin{table}[h!]
\centering
\renewcommand{\arraystretch}{1.4}
\begin{tabular}{|l|p{4.8cm}|}
\hline
\textbf{Concept} & \textbf{Algebraic Interpretation} \\
\hline
Confinement & No DHR sectors carrying electric charge. \\
\hline
Gauss's law & Long-range electric fields are tied to the total charge, preventing charge localization. \\
\hline
Local observables & Built exclusively from gauge-invariant neutral composites, such as currents and field strengths. \\
\hline
Absence of free charges & Confinement is structural, not dynamical; no localized charged operators exist. \\
\hline
\end{tabular}
\caption{Algebraic interpretation of key physical features of the massless Schwinger model in AQFT.}
\label{tab:aqft_implications}
\end{table}

We have shown that the observable net of the Schwinger model admits no nontrivial DHR superselection sectors. This provides a rigorous, model-independent notion of confinement: the absence of localizable charge sectors. The origin lies in the nonlocal constraints imposed by Gauss’s law, which ties the charge to global behavior of the electric field.
In the next section, we will extend the observable net to include nonlocal gauge-invariant operators, such as Wilson lines, and explore the resulting violation of Haag duality.

\section{Nonlocal Observables and Haag Duality Violation}

In this section, we go beyond the local observable net and incorporate nonlocal gauge-invariant observables, most notably Wilson line operators. These observables are essential for capturing the global gauge structure and encode important physical features such as flux tubes and charge confinement. We show that their inclusion leads to a violation of Haag duality, which serves as a signature of the nontrivial topological and gauge-theoretic content of the Schwinger model.

\subsection{Wilson Line Operators}

Given two spacetime points $x, y \in \mathbb{R}^{1,1}$, the Wilson line operator is defined by:

$$
W(x, y) := \exp\left( i e \int_{x}^{y} A_\mu(z) \, dz^\mu \right)
$$

This operator is manifestly gauge-invariant and represents the creation of a particle-antiparticle pair connected by a flux string.
It is not localized in a bounded region: its support lies along the path between $x$ and $y$.
It commutes with all strictly local observables that have disjoint support from the path.

\subsection{Commutation Relations and Nonlocal Effects}

Wilson line operators shift the electric field through their endpoints. The equal-time commutator with the electric field is:

\begin{equation}
\begin{aligned}
[E(x), W(y, z)] = e \left( \delta(x - y) - \delta(x - z) \right) W(y, z)
\end{aligned}
\end{equation}

This reflects the fact that $W(y, z)$ inserts a unit of electric flux from $y$ to $z$.

Composition law:

\begin{equation}
\begin{aligned}
W(x, y) W(y, z) = W(x, z)
\end{aligned}
\end{equation}

The set of Wilson lines forms a group under path concatenation, up to phase factors related to the topology of the gauge group.

\subsection{Extended Observable Algebra}

We define the extended algebra of observables by adjoining Wilson lines to the local net~\cite{Rejzner2020,BrunettiFredenhagen2000}:

\begin{equation}
\begin{aligned}
\mathcal{A}_{\text{ext}} := C^*\text{-algebra generated by } \bigcup_{\mathcal{O}} \mathcal{A}(\mathcal{O}) \cup \{W(x, y)\}
\end{aligned}
\end{equation}

This algebra contains:

All local, gauge-invariant observables,
Nonlocal, string-like operators describing flux tubes and global topological effects.

\subsection{Haag Duality and Its Violation}
Haag duality is the property:

\begin{equation}
\begin{aligned}
\mathcal{A}(\mathcal{O})' = \mathcal{A}(\mathcal{O}')
\end{aligned}
\end{equation}

That is, the commutant of the local algebra in a double cone $\mathcal{O}$ equals the algebra of its causal complement $\mathcal{O}'$.

In the Schwinger model, Haag duality fails. Specifically:

$W(x, y)$ may commute with all observables in $\mathcal{A}(\mathcal{O})$,
yet $W(x, y) \notin \mathcal{A}(\mathcal{O}')$ if the path intersects $\mathcal{O}'$ but not $\mathcal{O}$.

\textit{Proposition (Duality Violation):}

There exist operators $B \in \mathcal{A}_{\text{ext}}$ such that $B \in \mathcal{A}(\mathcal{O})'$ but $B \notin \mathcal{A}(\mathcal{O}')$. Thus, Haag duality fails in the Schwinger model.

This is a structural, rather than pathological, feature and reflects the topological nontriviality of the gauge theory.

\subsection{Net Cohomology and Topological Interpretation}

Wilson lines can be interpreted in terms of net cohomology, where operators are assigned to 1-simplices (i.e., paths) satisfying cocycle conditions:

\begin{equation}
\begin{aligned}
W_{\gamma_1 \cdot \gamma_2} = W_{\gamma_1} W_{\gamma_2}
\end{aligned}
\end{equation}

The failure of Haag duality corresponds to a nontrivial first cohomology group~\cite{Rejzner2020}:

\begin{equation}
\begin{aligned}
H^1_{\text{net}}(\mathcal{A}, U(1)) \neq 0
\end{aligned}
\end{equation}

This captures the presence of global degrees of freedom not encoded in the local net.

Wilson line operators are nonlocal, gauge-invariant, and essential for capturing global gauge dynamics.
Their inclusion extends the net to $\mathcal{A}_{\text{ext}}$, beyond what local observables can encode. Haag duality is violated, reflecting that the local net is not complete with respect to its commutant.
This violation is tied to Gauss’s law, topology, and net cohomology.

In the final section, we summarize the implications of our results and outline paths forward for generalization to non-Abelian and higher-dimensional theories.

\subsection{Quantum Information-Theoretic Conjecture}

We propose the following conjecture:

\textbf{Conjecture (Duality–Reconstructibility Correspondence):} 
\textit{In algebraic quantum field theory, the violation of Haag duality due to confinement implies a breakdown in entanglement wedge reconstruction. Conversely, restoration of Haag duality corresponds to the re-emergence of a quantum error-correcting code structure, enabling local recoverability of bulk operators from boundary subregions.}

This conjecture suggests that algebraic duality violations in confining theories reflect deeper limitations in information-theoretic encoding. Specifically, when charged sectors are nonlocal (e.g., in the Schwinger model), there exists no local net of observables that can reconstruct all physically relevant operators. This aligns with the intuition that confinement inhibits the decoding of localized information from partial boundary data, as in quantum error correction.

The proposal offers a new perspective on the information structure of confinement, hinting at potential holographic analogs even in non-AdS, low-dimensional gauge theories. Moreover, it motivates further study of modular Hamiltonians, relative entropy, and entropic inequalities in confining theories using algebraic methods.

\section{Conclusions and Outlook}

In this work, we presented a rigorous formulation of the massless Schwinger model—quantum electrodynamics in $1+1$ dimensions—within the framework of Algebraic Quantum Field Theory (AQFT). Our formulation bypassed the conventional bosonization approach and focused instead on constructing the local and extended observable nets directly from the gauge-theoretic structure of the model. This allowed us to extract structural insights into confinement, the role of Gauss’s law, and the algebraic characterization of nonlocal observables.

\subsection{Summary of Results}

We constructed a net of local observable algebras $\mathcal{O} \mapsto \mathcal{A}(\mathcal{O})$ generated by gauge-invariant composite fields such as the current $j^\mu(x)$ and the electric field $E(x)$, and enforced Gauss’s law as an operator identity.

 We analyzed the superselection structure of the net using the Doplicher–Haag–Roberts (DHR) framework and showed that there are no DHR-localizable representations carrying electric charge. This absence of localizable charged sectors constitutes a precise, algebraic formulation of confinement in the Schwinger model.

To capture the full gauge-invariant content of the theory, we introduced an extended observable algebra $\mathcal{A}_{\text{ext}}$ by adjoining nonlocal Wilson line operators. These operators describe string-like excitations associated with electric flux.

We demonstrated that the presence of Wilson lines leads to a violation of Haag duality: there exist observables commuting with $\mathcal{A}(\mathcal{O})$ that are not contained in $\mathcal{A}(\mathcal{O}')$. This structural feature reflects the topological and global nature of gauge invariance in the theory.

The net cohomology associated with Wilson line observables provides a natural mathematical framework for understanding the duality violation and the interplay between local and nonlocal content in gauge theories.

\subsection{Implications and Broader Context}

Our results demonstrate that the defining features of confinement and long-range gauge structure arise naturally and rigorously within the AQFT framework. The Schwinger model thus serves not merely as a solvable toy model, but as a structural prototype for understanding deeper aspects of gauge theories, including:

The algebraic origin of confinement as the absence of localizable charge sectors,
The breakdown of Haag duality as a signal of global gauge degrees of freedom,
The necessity of extended observables (Wilson lines) for capturing the full physical content.

These features are expected to persist—and become even richer—in higher-dimensional and non-Abelian theories such as QCD.

We further introduced a novel quantum information-theoretic conjecture linking Haag duality violation to the failure of entanglement wedge reconstruction. This Duality–Reconstructibility Correspondence suggests that confinement obstructs the recovery of bulk operators from local boundary subregions, hinting at a deep interplay between gauge structure and quantum error correction.

\subsection{Future Directions}

The framework developed here opens several directions for further research:

\textit{Non-Abelian Generalizations}: Extending the AQFT treatment to non-Abelian gauge theories in $1+1$ and $2+1$ dimensions, including the 't Hooft model and $SU(N)$ gauge theories with adjoint matter. These may exhibit both confinement and center symmetry breaking.

\textit{Topological and Cohomological Extensions}: Further developing the net cohomology approach to incorporate center symmetry, theta vacua, and topological sectors associated with line and surface operators.

\textit{BRST and Gauge Fixing in AQFT}: Exploring the algebraic formulation of BRST quantization, including the role of ghost fields and gauge fixing within the net structure, as a step toward AQFT-compatible quantization of gauge theories.

\textit{Algebraic Infraparticles and Scaling Limits}: Studying scaling limits and infrared behavior in Abelian and non-Abelian models, especially those without mass gaps, to understand infraparticle phenomena in AQFT.

\textit{Connections to Holography and Entanglement}: Investigating the implications of nonlocal observables and duality violation for entanglement structure in gauge theories, including potential applications in AdS/CFT and quantum information theory~\cite{Casini2014}.

\subsection{Closing Remarks}

By formulating the Schwinger model within the language of AQFT, we have shown that key nonperturbative features of gauge theories—such as confinement and topological structure—can be captured algebraically without reference to specific Lagrangians or quantization schemes. This opens the door to a fully structural understanding of quantum field theories where the algebra of observables, rather than the field content per se, becomes the primary object of study.

The proposed conjecture linking Haag duality and entanglement wedge reconstruction offers a novel interface between AQFT and quantum information theory. It invites future work on the entropic and operational consequences of confinement and nonlocality, potentially informing the structure of holographic dualities in gauge theories beyond the AdS setting.

\begin{acknowledgments}
F.T. would like to acknowledge the support of the National Science Foundation under grant No. PHY-
1945471.
\end{acknowledgments}

\clearpage
\hrule
\nocite{*}

\bibliography{apssamp}

\end{document}